\documentclass[twocolumn,twoside,slac_two]{revtex4}

\usepackage{graphicx}
\usepackage{fancyhdr}
\newcommand{\etal}{et al.}
\newcommand{\Hspls}{\ensuremath{H^2_+(\qsq)}}
\newcommand{\Hsmin}{\ensuremath{H^2_-(\qsq)}}
\newcommand{\Hszer}{\ensuremath{H^2_0(\qsq)}}
\newcommand{\Hpls}{\ensuremath{H_+(\qsq)}}
\newcommand{\Hmin}{\ensuremath{H_-(\qsq)}}
\newcommand{\Hzer}{\ensuremath{H_0(\qsq)}}
\newcommand{\hzer}{\ensuremath{h_0(\qsq)}}

\newcommand{\Hint}{\ensuremath{\hzer\,\Hzer}}

\newcommand{\krzb}{\ensuremath{\overline{K}^{*0}}}
\newcommand{\krzmndk}{\ensuremath{D^+ \rightarrow \krzb \mu^+ \nu}}
\newcommand{\krzlndk}{\ensuremath{D^+ \rightarrow \krzb \ell^+ \nu}}
\newcommand{\philndk}{\ensuremath{D_s^+ \rightarrow \phi\; \ell^+ \nu}}

\newcommand{\kpilndk}{\ensuremath{D^+ \rightarrow K^- \pi^+ \ell^+ \nu }}

\newcommand{\thv}{\ensuremath{\theta_\textrm{v}}}
\newcommand{\thl}{\ensuremath{\theta_\ell}}
\newcommand{\costhv}{\ensuremath{\cos\thv}}

\newcommand{\costhl}{\ensuremath{\cos\thl}}

\newcommand{\sinthlsq}{\ensuremath{\sin^2\thl}}
\newcommand{\qsq}{\ensuremath{q^2}}

\newcommand{\bw}{\ensuremath{\textrm{BW}}}
\newcommand{\mkpi}{\ensuremath{m_{K\pi}}}

\newcommand{\rtwo}{\ensuremath{r_2}}

\newcommand{\rvee}{\ensuremath{r_v}}
\newcommand{\mysection}[1]{\section{#1}}

\newcounter{saveeqn}%
\begin{document}

\title{Recent results on fully leptonic and semileptonic charm decays}

%

\author{Jim Wiss}
\affiliation{University of Illinois, 1110 W. Green, Urbana IL, 61801}

\begin{abstract}
We begin with giving some motivation for the study of charm semileptonic and fully leptonic
decays. We turn next to a discussion of semileptonic absolution branching fraction results
form CLEO-c.  Two exciting high statistics results on fully leptonic decays of the $D^+ \rightarrow \mu^+ \nu$ and
$D_s^+ \rightarrow \mu^+ \nu$ from CLEO-c and BaBar are reviewed. We turn next to a discussion of recent results
on charm meson decay to pseudo-scalar $\ell nu$ decays from FOCUS, BaBar, and CLEO-c.  We conclude with a review of 
charm meson decay into Vector $\ell \nu$ .
\end{abstract}
\maketitle

\thispagestyle{fancy}


\section{Introduction}
\begin{figure}[tbph!]
 \begin{center}
  \includegraphics[width=3.in]{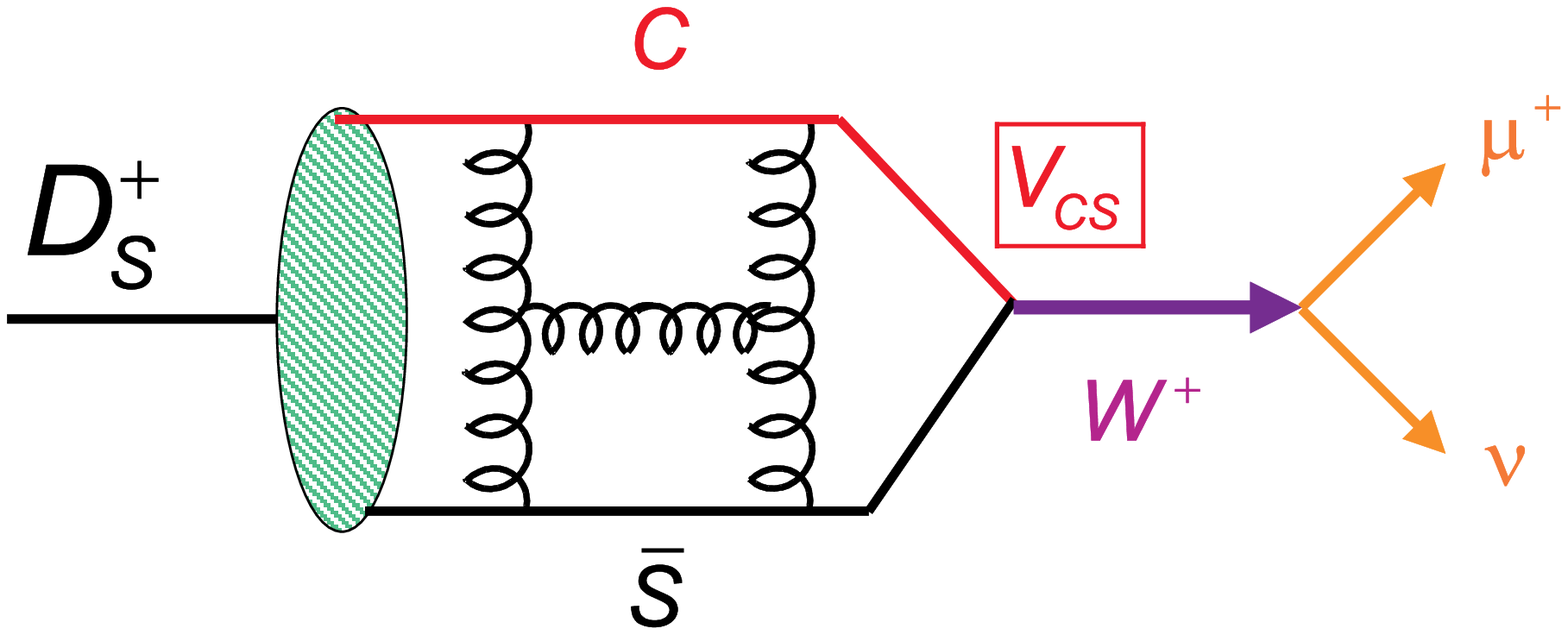}
  \includegraphics[width=3.5in]{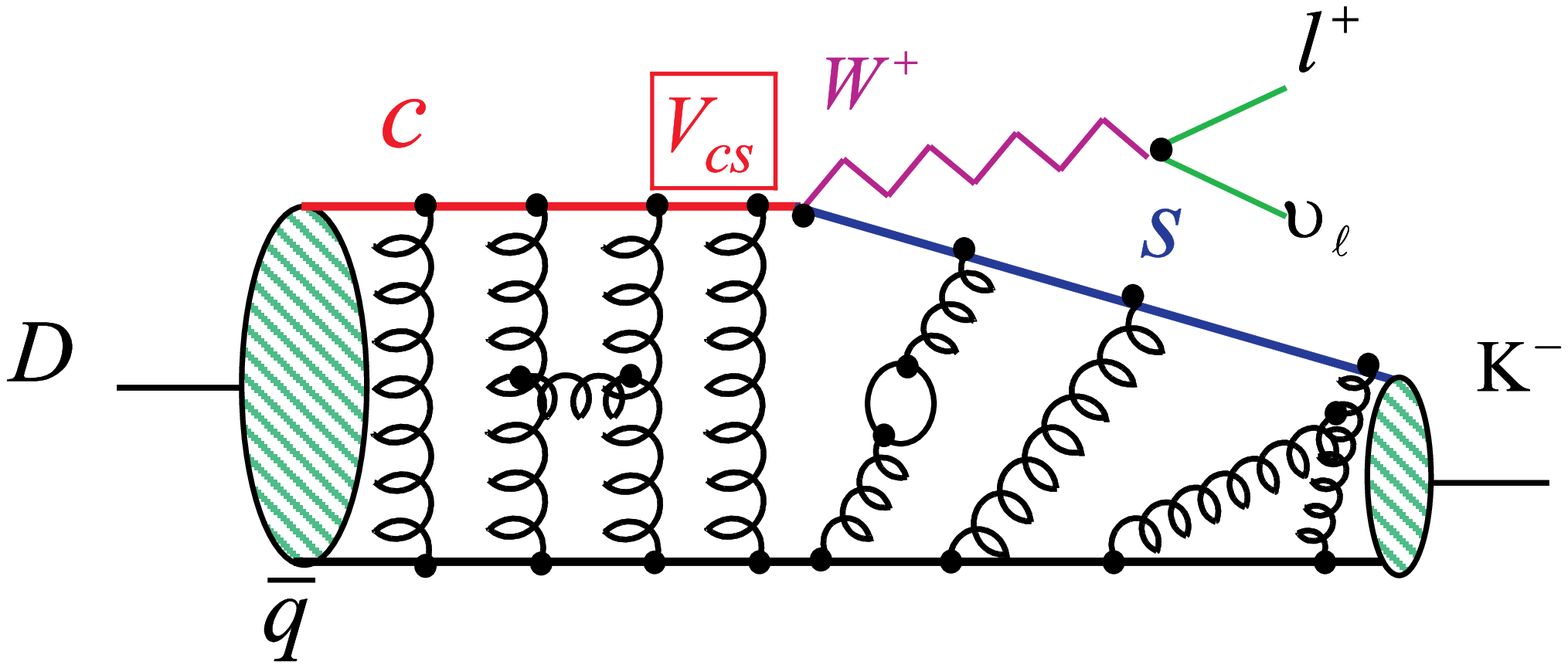}
  \caption{Diagrams for the fully leptonic (top) and semileptonic (bottom) decay of charmed mesons. The
QCD complications are contained in a decay constant $f_D$ for the fully leptonic case, and
 \qsq{} dependent form factors for semileptonic decays. Both processes, in principle, provide measurements of 
CKM matrix elements 
\label{cartoon}} \end{center}
\end{figure}

Figure \ref{cartoon} shows cartoons of the $D_s^+ \rightarrow \ell^+ \nu$ fully leptonic
process and the $D^0 \rightarrow K^- \ell^+ \nu$ semileptonic decay process.  
All of the  hadronic complications for this process are contained in the decay constant for fully 
leptonic decay or the $\qsq{}$ dependent form factor for semileptonic processes. Both are computable 
using non-perturbative methods such as LQCD.  Although both processes can in principle provide a  determination of charm CKM elements, one frequently uses the (unitarity constrained) CKM measurements, lifetime, and branching
fraction to measure the $f_D$ decay constant or the \qsq{} integral of the square of the semileptonic form
factor.  These can then be compared to LQCD predictions to provide an incisive test of this technique.  The $\qsq{}$ dependence of the semileptonic form factor can also be directly measured
and compared to theoretical predictions.  

The hope is that charm semileptonic and fully leptonic decays can provide high statistics, precise tests of LQCD calculations and thus
validate the computational techniques for charm.  
Once validated, the same LQCD techniques can be used in related calculations for $B$-decay
and thus produce CKM parameters with significantly reduced theory systematics.  For example the recent $B_s^0$ mixing rate measurement
by CDF and D0 is proportional to the squared $f_{B_s}$ decay constant.  The analogous  $f_{D_s}$ computed using similar
methods was recently measured by the BaBar Collaboration. 

\begin{figure}[tbph!]
 \begin{center}
  \includegraphics[width=2.in]{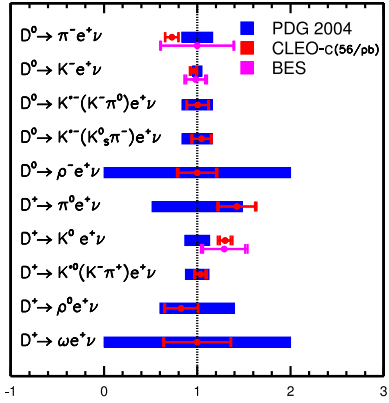}
  \caption{ Ten exclusive absolute semileptonic branching fractions measured by the CLEO collaboration divided by 
the previous world average values compiled by PDG2004.  The CLEO fractional errors are shown with red error bars within
the blue boxes. The PDG2004 relative 
errors are shown as blue boxes.  Even though these results are based on
only about 20\% of CLEO's present sample, in most cases they are more precise than the previous world average.  
\label{exclusive}} \end{center}
\end{figure}

\mysection{Absolute Semileptonic Decay Branching Fractions from CLEO}
These results are based on the first 56 pb$^{-1}$ of CLEO charm threshold running at the $\psi(3770)$  At this energy
charm is produced by either in $D^0 \bar D^0$ or $D^+ D^-$ final states since there is 
not enough kinematic room to produce an additional pion~\cite{dtag}.  This is a particularly desirable
environment for measuring absolute branching fractions since one essentially divides the corrected number of observed $D X$
events where $X$ decays into a given semileptonic state to the total corrected number of states with a reconstructed D. The semileptonic branching fraction results \cite{cleobr} are summarized in Fig. \ref{exclusive}.

\begin{figure}[tbph!]
 \begin{center}
  \includegraphics[width=3.in]{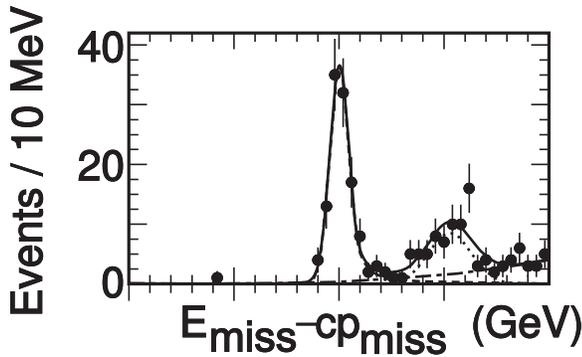}
  \caption{ 
Illustration of a $\pi^+ e^+ \nu$ signal obtained in the 
first 57 pb$^{-1}$ of CLEO-c running at the $\psi(3770)$.  The signal forms a peak near
$U \equiv E_{\rm Miss} - c |\vec P_{\rm Miss}| \approx 0$. A misidentification 
background from  $D^0 \rightarrow K^- e^+ \nu$ is well displaced from the signal. \label{pienu}} \end{center}
\end{figure}

Figure \ref{pienu} gives an example of the CLEO $D^0 \rightarrow \pi^- e^+ \nu$ signal.  The signal
appears as a peak centered at zero in the variable $U \equiv E_{\rm Miss} - c |\vec P_{\rm Miss}|$ in 
events with a fully reconstructed ${\bar D}^0$.  To get a contribution to the signal peak, one must have
a single missing $\nu$ and the proper masses must be assigned to the charged semileptonic decay daughters in constructing $E_{\rm Miss}$. 
Figure \ref{pienu} illustrates the power of this kinematic constraint by showing the separation
between the $D^0 \rightarrow \pi^- e^+ \nu$ signal and a $D^0 \rightarrow K^- e^+ \nu$ background where
the $K^-$ has been misidentified a $\pi^-$ by the particle identification system.   Even though
the $K^- e^+ \nu$ dominates over $\pi^- e^+ \nu$  by about an order of magnitude, its 
contamination near $U \approx 0$ is very small and manageable.

\begin{figure}[tbph!]
 \begin{center}
  \includegraphics[width=2.in]{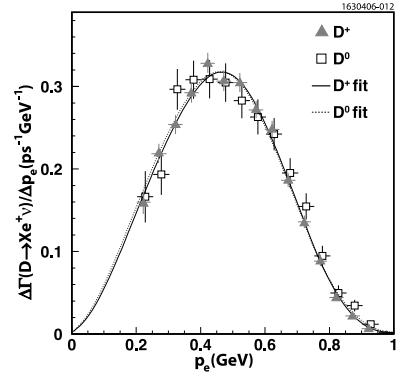}
  \caption{ The momentum spectrum of electrons/positrons for the CLEO-c inclusive semileptonic
selection for $D^0$ and $D^+$ candidates. The fitted curves, based a Monte Carlo model, are used to extrapolate the spectrum below the electron momentum cut-off of 200 MeV$/c$.   
\label{spectrum}} \end{center}
\end{figure}

CLEO has also reported on inclusive semileptonic decays of the $D^0$ and the $D^+$. The 
data are based 281 pb$^{-1}$ of their $\psi(3770)$
running.  Figure \ref{spectrum} compares the electronic momentum spectrum obtained against tagged
$D^0$ and the $D^+$ along the curves used to extrapolate the spectrum below their
cut-off of $ P_e < ~200~MeV/c$. Roughly 7.6\% of the semileptonic decays produce
an electron below this $200~MeV$ electron momentum cut according to their Monte Carlo model generated with ISGW2 form factors.
A 1\% systematic uncertainty is assessed on the inclusive $\mathcal{B}$ values summarized in Table \ref{IncBR} for the momentum extrapolation below 200 MeV/$c$.

The table summarizes the preliminary results for the $D^0$ and $D^+$ inclusive semileptonic branching fractions and compares each result to the 
sum of the CLEO exclusive mode branching fractions. The known exclusive modes come close to saturating the inclusive modes
although there might be some room for additional, unmeasured exclusive states.

\begin{table}[htp]
\caption{Inclusive semileptonic branching fractions compared to the sum of the semi exclusive branching fractions 
measured by CLEO.\label{IncBR}}
\begin{center}
\begin{tabular}{|l|c|} \hline
$D^0 \rightarrow X e^+ \nu$ & (6.46 $\pm$ 0.17 $\pm$ 0.13)\% \\ 
$\Sigma_i  \mathcal{B}_i\left(D^0 \rightarrow X e^+ \nu \right)$ &(6.1 $\pm$ 0.2 $\pm$ 0.2)\% \\ \hline 
 $D^+ \rightarrow X e^+ \nu$ & (16.13 $\pm$ 0.20 $\pm$ 0.33)\% \\ 
 $\Sigma_i  \mathcal{B}_i \left(D^+\rightarrow X e^+ \nu \right)$& (15.1 $\pm$ 0.5 $\pm$ 0.5)\% \\ \hline
\end{tabular}
\end{center}
\end{table}
One can also use the ratio of the exclusive $D^+$ and $D^0$ semileptonic branching fractions and the known $D^0$ and $D^+$
lifetimes to measure the ratio of $D^+$ and $D^0$ semileptonic widths. 

\begin{eqnarray}
{\frac{{\Gamma _{D^ +  }^{SL} }}{{\Gamma _{D^0 }^{SL} }} = \frac{{B_{D^ +  }^{SL} }}{{B_{D^0 }^{SL} }} \times \frac{{\tau _{D^0 } }}{{\tau _{D^ +  } }}\; = 0.985 \pm 0.028 \pm 0.015}
\end{eqnarray}

The value is consistent with unity as expected from isospin symmetry. The errors on the new 
CLEO width ratio represents a considerable improvement over previous data.

\mysection{Fully leptonic decays from CLEO and BaBar}

Charm fully leptonic decays  are difficult to study because of their very low branching ratios. The rate is low since the charged lepton is forced into an unnatural helicity state to conserve angular momentum. The decay width
is proportional to two powers of the lepton mass -- in this case the mass of the muon.   
\begin{eqnarray}
B(D^{}  \to \mu \nu )/\tau _{D^{} }  = \frac{{G_F^2 }}{{8\pi }}f_{D^{} }^2 m_\mu ^2 M_{D^{} } \left( {1 - \frac{{m_\mu ^2 }}{{M_{D^{} }^2 }}} \right)\left| {V_{cq} } \right|^2 {\rm{ }}
\end{eqnarray}
For example CLEO \cite{fdpls} measures $\mathcal{B}\left(D^+ \rightarrow \mu^+ \nu \right) = (4.40 \pm 0.66 \pm 0.1)\times 10^{-4}$ , while BaBar has a preliminary
measurement of $\mathcal{B}\left(D_s^+ \rightarrow \mu^+ \nu\right) = (6.5 \pm 0.8 \pm 0.3 \pm 0.9)\times 10^{-3}$. The order of magnitude larger $D_s^+ \rightarrow \mu^+ \nu$ fully leptonic $\mathcal{B}$ reflects the factor-of-two difference in the $D_s^+$ and $D^+$ lifetimes as well as the fact
that the $D_s^+$ fully leptonic decay is a Cabibbo favored process where $D^+ \rightarrow \mu^+ \nu$ is not.

\begin{figure}[tbph!]
\begin{center}
  \includegraphics[width=2.in]{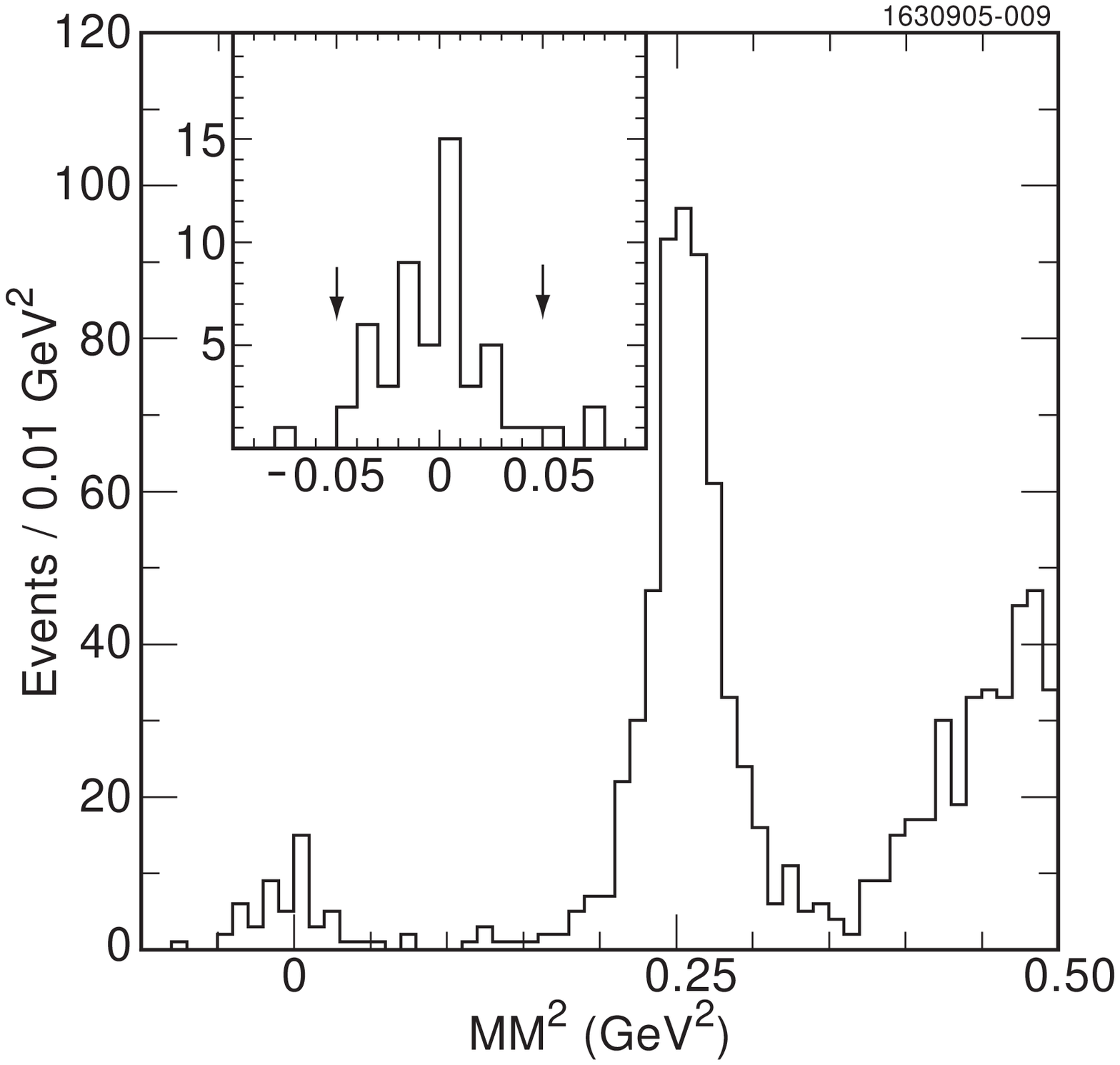}
\includegraphics[width=3.in]{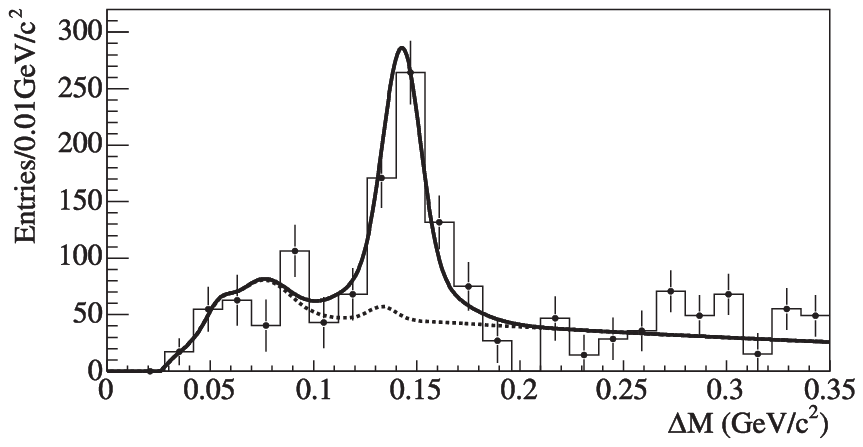} 
\caption{Signals for $D^+ \rightarrow \mu^+ \nu$ from CLEO (top) and 
$D_s^+ \rightarrow \mu^+ \nu$ from BaBar (bottom). The $D^+ \rightarrow \mu^+ \nu$ signal
appears as a peak in the missing mass distributions in CLEO tagged events where
all tracks are reconstructed but the neutrino.  The $D_s^+ \rightarrow \mu^+ \nu$ signal
forms a peak in the $D_s^{+*} \rightarrow \gamma D_s^+$ mass difference plot.
\label{FD}}
\end{center}
\end{figure}

Both the CLEO signal for $D^+ \rightarrow \mu^+ \nu$ ($\approx 50$ signal events) and the preliminary BaBar signal for $D_s^+ \rightarrow \mu^+ \nu$
($489 \pm 55$ signal events) are shown in Fig. \ref{FD}.  

The CLEO signal, based  281 pb$^{-1}$ of the $\psi(3770)$ running, is a peak near zero in the missing-mass in events where there is a single track recoiling against a fully reconstructed $D^-$. 
The prominent peak centered to the right of the neutrino peak, near a missing mass
of $\approx 0.25~\rm{GeV}^2/c^2$, presumably corresponds to charm decays with an unreconstructed $K_L$. 

The BaBar analysis is very 
different since they are running at the $\Upsilon (4S)$ which is far from charm threshold.  They observe the $D_s^+ \rightarrow \mu^+ \nu$
decay by observing a peak in the $\Delta m \equiv m(\mu^+ \nu ~\gamma)-m(\mu^+ \nu)$ mass difference corresponding to the 
decay $D_s^{+*} \rightarrow \gamma D_s^{+}$. The neutrino four-vector is estimated from the missing momentum in the event along with the application
of a $D_s^+$ mass constraint when the neutrino is combined with the reconstructed muon.    There is a slight peaking background near $\approx 0.07 ~\rm{GeV}/c$
due to photons originating from $\pi^0$ rather than $D_s^{+*}$ decay. The dashed background is due to 
$D_s^{+*} \rightarrow \gamma D_s^{+}\rightarrow \gamma (\tau^- \nu)$ decays. It is interesting to note that although the $D^{+*} \rightarrow \gamma D^{+}\rightarrow \gamma (\mu^+ \nu)$ background will peak at essentially the same $\Delta m$ as the $D_s^{+*} \rightarrow \gamma D_s^{+}\rightarrow \gamma (\mu^+ \nu)$ signal, this background will be essentially negligible in light of the order of magnitude lower  $\mathcal{B}\left(D^+ \rightarrow \mu^+ \nu \right)$ and the fact that $\mathcal{B}\left(D_s^{+*} \rightarrow \gamma D_s^{+}\right)$ is about 60 $\times$ larger than $\mathcal{B}\left(D^{+*} \rightarrow \gamma D^{+}\right)$.

\begin{table}[htp]
\caption{Fully leptonic decay constants\label{FDT}}
\begin{center}
\begin{tabular}{|l|c|} \hline
$f_{D+}$ LQCD (FNAL/MILC)\cite{milc} & 201 $\pm$ 3 $\pm$17 MeV \\\hline
$f_{D+}$ (CLEO)\cite{fdpls} & 222.6$\pm$16.7$^{+2.8}_{-3.4}$ MeV \\  
$f_{Ds}$ (BaBar) & 279 $\pm$ 17 $\pm$ 6 $\pm$19 MeV \\ \hline 
$f_{Ds}$/ $f_{D+}$ BaBar/CLEO & 1.25 $\pm$ 0.14 \\ \hline
\end{tabular}
\end{center}
\end{table}

The CLEO $f_{D+}$ result is consistent with the latest LQCD estimate from the FNAL/MILC collaboration\cite{milc} and has comparable errors.
The BaBar $f_{Ds}$ result is about 25 \% higher than $f_{D+}$ as expected in LQCD calculations. Both $f_D$ decay constants are measured to about 8 \%. The major systematic error for the BaBar measurement is $\pm$ 19 MeV, due to 
the $13\%$ uncertainly in the $\mathcal{B}\left(D_s^+ \rightarrow \phi \pi^+\right)$ measured by BaBar which was used to normalize their 
$D_s^+ \rightarrow \mu^+ \nu$ signal.

\mysection{ Pseudoscalar $\mathbf{\ell\nu}$ decays from FOCUS, BaBar, Belle, and CLEO}

The below equation gives the expression for the differential decay width for $D \to P\ell \nu$ where P is a pseudoscalar meson.
\begin{eqnarray}
\frac{{d\Gamma \left( {D \to P\ell \nu } \right)}}{{dq^2 }} = \frac{{G_F^2 \left| {V_{cq} } \right|^2 P_P^3 }}{{24\pi ^3 }}\left\{ {\left| {f_ +  (q^2 )} \right|^2  + O(m_l^2 )} \right\}
\end{eqnarray}
The pseudoscalar semileptonic decay -- in the limit of low charged lepton mass -- is controlled by a single form factor $f_+ (\qsq{})$. An important motivation for studying pseudoscalar semileptonic decays is to compare the measured $f_+ (\qsq{})$ to the calculated
$f_+ (\qsq{})$ using techniques such as LQCD. The $P_P^3$ factor (where $P_P$ is the momentum of the pseudoscalar in the $D$ rest frame) creates a strong
peaking of $d \Gamma/d \qsq{}$ at low \qsq{}. Unfortunately the low \qsq{} region is where discrimination between different $f_+ (\qsq{})$ models is the poorest, and LQCD calculations are the most difficult. To the extent that $f_+(\qsq{})$ calculations are trusted, a measurement of the pseudoscalar semileptonic decay widths can provide new measurements of the CKM matrix elements.

\begin{figure}[tbph!]
\begin{center}
  \includegraphics[width=2.5in]{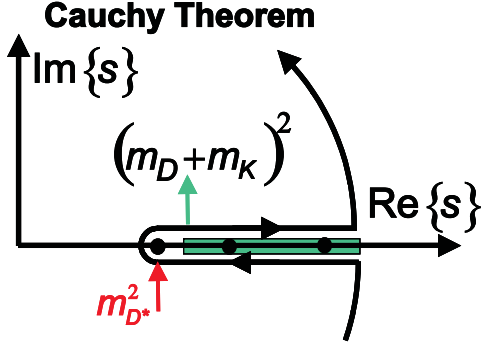}
  \caption{ The form factor is assumed to be an analytic function with pole singularities at the 
masses of bound states, and cuts that start at the start of the continuum.  One can use Cauchy's
theorem with the indicated contour to write an dispersion expression for $f_+ (\qsq{})$ in the physical range $0 < \qsq{} < \left(m_D - m_K \right)^2$. 
\label{cut}} 
\end{center}
\end{figure}

We begin discuss studies on the \emph{shape} of $f_+ (\qsq{})$ for the decay $D^0 \rightarrow K^- \ell^+ \nu$. An early parameterization for $f_+ (\qsq{})$ used spectroscopic pole dominance.  This is 
based on a dispersion relation obtained using Cauchy's Theorem under the assumption that $f_+(\qsq{})$
is an analytic, complex function as illustrated in Fig. \ref{cut} for $D^0 \rightarrow K^- e^+ \nu$.
The $f_+(\qsq{})$ singularities will consist of simple poles at the $D^0 K^+$ vector bound states (e.g. $D_s^{*+}$) and
cuts beginning at the $D^0 K^+$ continuum ($\qsq{} > (M_D + M_K)^2$). The dispersion relation
gives $f_+(\qsq{})$ as a sum of the spectroscopic pole and an integral over the cut:
\begin{eqnarray}
f_ +  (q^2 ) = \frac{\mathcal{R}}{{m_{D_s^*}^2  - q^2 }} + \frac{1}{\pi }\int_{\left( {m_D  + K} 
\right)^2 }^\infty  {\frac{{{\mathop{\rm Im}\nolimits} \left\{ {f_ +  (s)} \right\}}}{{s - q^2  - 
i\varepsilon }}ds} 
\end{eqnarray}
Both the cuts and poles are beyond the physical $\qsq{}_{\rm max}$ and thus can never be realized. One might expect the spectroscopic pole to dominate as $\qsq{} \rightarrow m^2_{D_s*}$ 
as long as the pole were
well separated from the cut. Neither of these conditions is particularly well satisfied for $D^0 \rightarrow K^- e^+ \nu$.
The minimum separation from the pole is $\sqrt{\qsq{}_{\rm max}} - m_{D_s*} = 0.74~\rm{GeV}$ which seems large on the scale of the charm system. The gap
between the pole and the start of the cut interval is only $0.25~\rm{GeV}$. Hence it does not appear that
the data ever gets ''close' to a ''well-isolated'' pole in  $D^0 \rightarrow K^- e^+ \nu$.

\begin{figure}[tbph!]
 \begin{center}
  \includegraphics[width=3.3in]{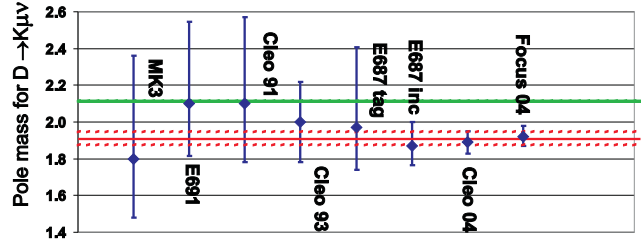}
  \caption{ Effective pole mass measurement in $D^0 \rightarrow K^- e^+ \nu$ over the years~\cite{mpole}.
The green line is the $m_{D_S^*}$ spectroscopic pole mass and is inconsistent with the 
average of the displayed data by 5.1 $\sigma$. \label{mpole}} \end{center}
\end{figure}

Several experiments have measured the ''effective'' pole mass in $D^0 \rightarrow K^- e^+ \nu$ decay over the years, 
where ${\rm{f}}_{\rm{ + }} (q^2 ) \propto 1/{(m_{pole}^2  - q^2)}$.  As Fig. \ref{mpole} shows, as errors have
improved over the years, it becomes clear that the effective pole is significantly lower than the spectroscopic 
pole, underscoring the importance of the cut integral contribution for this decay.

Becirevic and Kaidalov (1999) \cite{BK} proposed a new parameterization for $f_+(\qsq{})$ that would hopefully provide more insight
into the interplay between the spectroscopic pole and the cut integral contributions. 
\begin{eqnarray}
f_ +  (q^2 ) = \frac{{c_D m_{D*}^2 }}{{m_{D*}^2  - q^2 }} - \frac{{\alpha \gamma c_D m_{D*}^2 }}{{\gamma m_{D*}^2  - q^2 }}
\end{eqnarray}
Becirevic and Kaidalov represent the cut integral by an effective pole that is displaced from the spectroscopic pole
by a factor of $\sqrt{\gamma}$, and has a residue that differs from the spectroscopic pole by a factor $-\alpha$.  Becirevic and Kaidalov
use counting laws, and form factor relations in the heavy quark limit to argue that $\alpha= 1/\gamma$.  
This constraint
leads to a modified pole form with a single additional parameter $\alpha$ that describes the degree to which
the single spectroscopic pole fails to match $f_+(\qsq{})$ for a given process. 

\begin{eqnarray}
f_ +  (q^2 ) = \frac{{f_ +  (0)}}{{\left( {1 - q^2 /m_{D*}^2 } \right)\left( {1 - \alpha q^2 /m_{D*}^2 } \right)}}
\label{modpole}
\end{eqnarray}
The spectroscopic pole dominance limit is $\alpha \rightarrow 0$. As $\alpha$ is increased, the
effective cut pole both gets closer to $\qsq{}_{\rm max}$ limit while
simultaneously acquiring a stronger residue.  Both effects act
to create a faster $\qsq{}$ dependence than that of the spectroscopic pole thus creating an effective 
single effective pole with $m_{\rm pole} < m_{D_s*}$.  This is indeed what happens in the
data summarized in Fig. \ref{mpole}.

The Becirevic and Kaidalov parameterization has been used extensively in some of the calculational details of recent charm LQCD calculations of the $f_+$ form factor. The final $f_+(\qsq{})$
computed in reference \cite{milc} is well fit with a modified
pole form with $\alpha(K^- e^+ \nu) = 0.5 \pm 0.04$ and $\alpha(\pi^- e^+ \nu) = 0.44 \pm 0.04$.
It is interesting that the $\alpha$ parameters for the LQCD
calculations $D^0 \rightarrow \pi^- e^+ \nu$ are so similar to those for 
$D^0 \rightarrow K^- e^+ \nu$ given the different locations of their singularities.  For the $D^0 \rightarrow \pi^- e^+ \nu$, for example, $\qsq{}_{\rm max}$
lies much closer to the spectroscopic pole than the case in $D^0 \rightarrow K^- e^+ \nu$ and
for pion decay the $D^{*+}$ pole lies within the $D \pi$ continuum.

Although with present precision, data seems to match the modified pole form as well as the effective
pole form, it is not clear that the heavy quark limit really applies to charm semileptonic decay. Alternative $f_+(\qsq)$ parameterizations
have therefore been proposed in the literature \cite{Hill}.

\begin{figure}[tbph!]
 \begin{center}
  \includegraphics[width=1.5in]{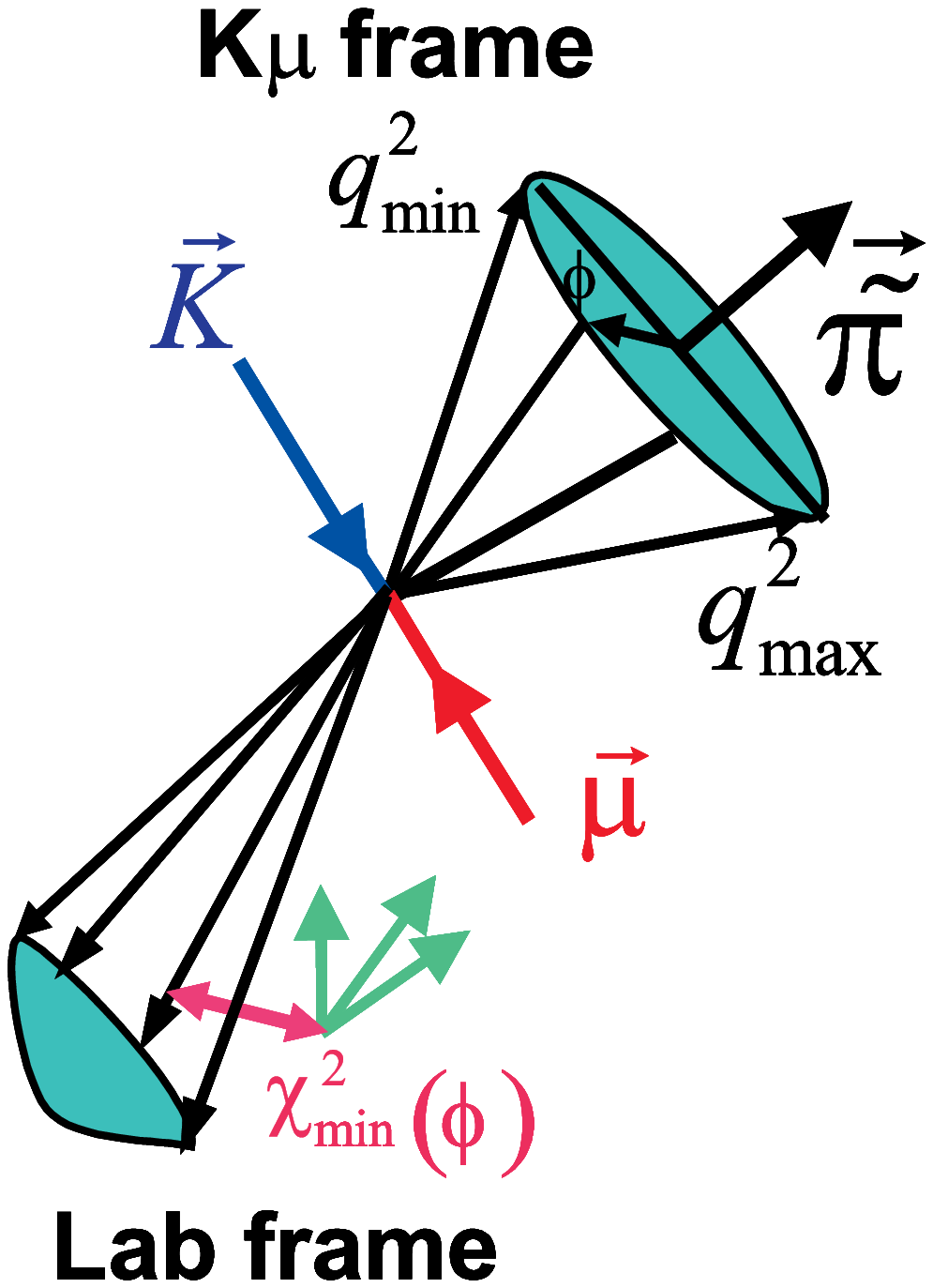}
  \includegraphics[width=2.5in]{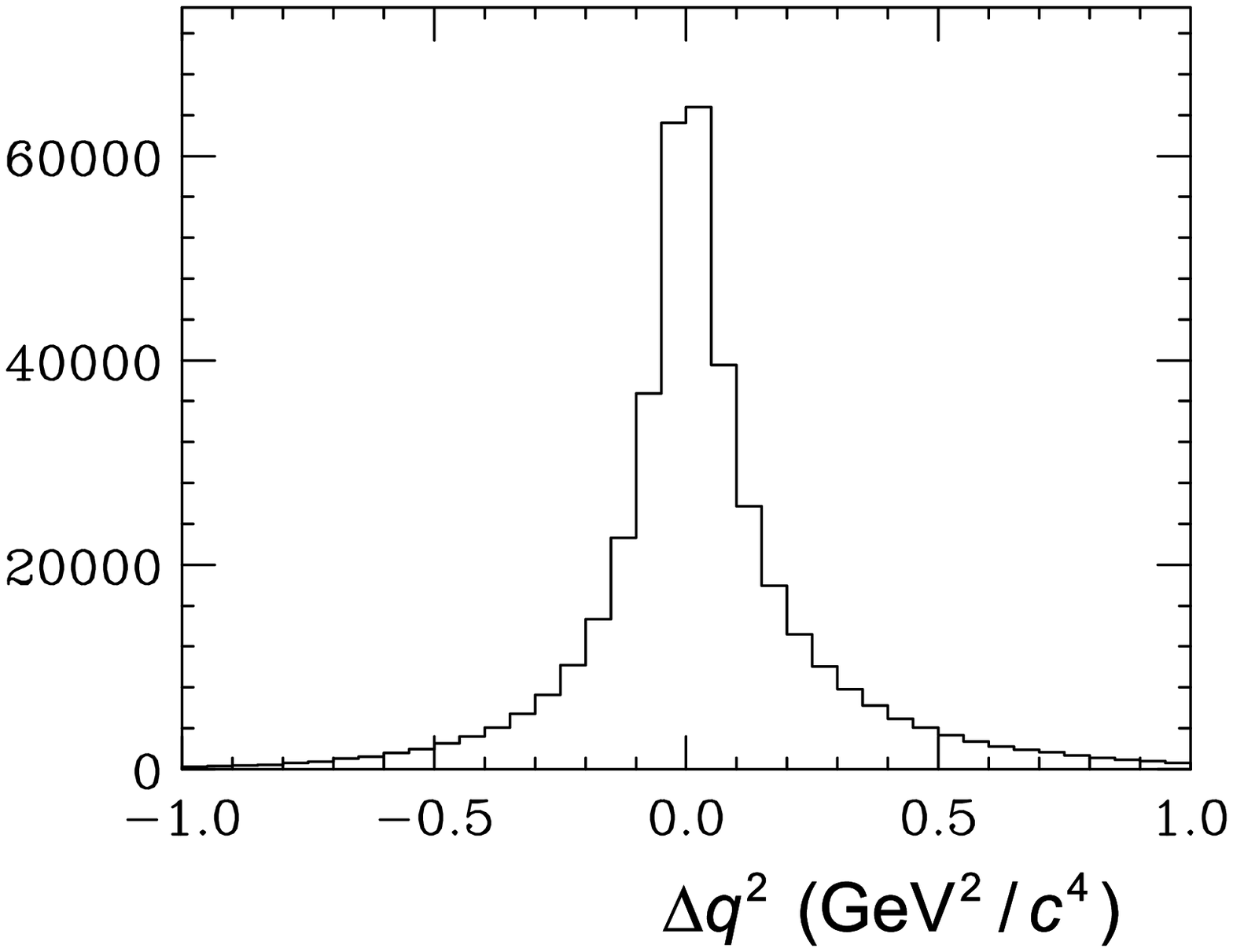}
  \caption{ Illustration of how the neutrino can be reconstructed in a fixed target
experiment such as FOCUS for the decay sequence $D^{*+} \rightarrow \tilde \pi^+ (K^- \mu^+ \nu)$ .
In the $K^- \mu^+$ rest frame the neutrino lies on a cone with a momentum and 1/2 angle 
given by the $D$ and $D^*$ mass constraint. One can the vary the azimuth along the cone to 
pick the solution where the $D$ passes closest to the primary vertex. On the right we show
the \qsq{} resolution obtained using this technique. \label{focus_closure}} \end{center}
\end{figure}

There are now several fine-bin, non-parametric measurements for $f_+(\qsq)$ for $D^0 \rightarrow K^- e^+ \nu$.  Essentially
the first of these was from FOCUS (2004).  The FOCUS data uses the decay chain
$D^{*+} \rightarrow \tilde \pi^+ (K^- \mu^+ \nu)$ and uses a signal consisting of $\approx 13000$ events after a tight $m(D^*) - m(D)$
cut.  Figure \ref{focus_closure} illustrates the method used by FOCUS to reconstruct the missing neutrino required to compute \qsq{}.
In the $K^- - \mu^+$ center of frame, the requirement that $K^-  \mu^+ \nu$ forms a $D^0$ determines the energy of the $\nu$. The
requirement that the $\tilde \pi^+ (K^-  \mu^+ \nu)$ forms a $D^{+*}$ restricts the neutrino to lie on a cone about the $\tilde \pi$
momentum. The $D^0$ momentum vector is directed against the neutrino in this frame. One then varies the azimuth about the neutrino
cone, boosts the $D^0$ momentum vector into the lab, and selects the azimuth where the $D^0$ comes closest to the primary
vertex in the photoproduced event.  The resultant \qsq{} resolution (also in Fig. \ref{focus_closure} )is roughly $\sigma (\qsq{}) \approx 0.20~\rm{GeV}^2$ which is comparable to the \qsq{} binning of 0.18 GeV$^2$.  A matrix based deconvolution technique is applied to the data.  The adjacent
$f_+(\qsq{})$ values have roughly a 65\% negative correlation owing to \qsq{} smearing between bins.  
 
\begin{figure}[tbph!]
 \begin{center}
  \includegraphics[width=3.in]{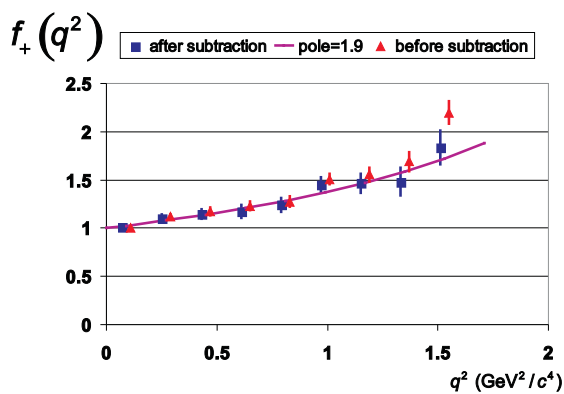}
  \caption{  $f_+(\qsq{})$ shapes in $K^- \ell+ \nu$ obtained by FOCUS \cite{focusqsq} prior
to charm background subtraction triangles and after background subtraction rectangles. The curve
is an effective pole fit with $m_{\rm pole} = 1.91~\rm{GeV}/c^2$. \label{fpls_focus}} \end{center}
\end{figure}

Figure \ref{fpls_focus} shows the deconvoluted $f_+(\qsq{})$ measurements both with and without subtraction of 
known charmed backgrounds.  It is intriguing to note that background only substantially affects the
highest \qsq{} bin. The curve shows an effective pole form with $m_{\rm pole} = 1.901 ~\rm{GeV}$
or a modified pole parameter of $\alpha = 0.32$. Both forms fit the data equally well.

\begin{figure}[tbph!]
 \begin{center}
  \includegraphics[width=3.in]{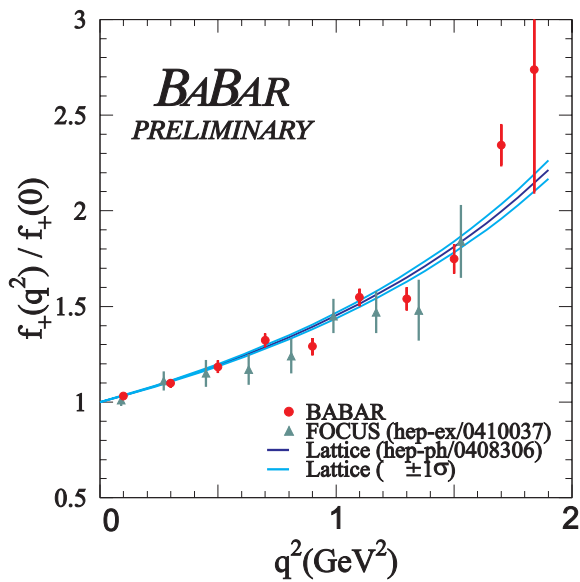}
  \caption{ Agreement between the FOCUS and BaBar $f_+(\qsq{})$ shapes in $K^- \ell^+ \nu$.
The results are also compared to a recent LQCD calculation\cite{milc} \label{fpls_babar}} \end{center}
\end{figure}

This year, the published results of CLEO III and FOCUS have joined by preliminary results from BaBar, CLEO-c, and Belle.
Figure \ref{fpls_babar} compares their measurement of the $f_+(\qsq)$ from $\approx 100K$ $D^0 \rightarrow K^- e^+ \nu$
to the FOCUS measurements and LQCD predictions~\cite{milc}. BaBar makes this measurement at the $\Upsilon (4S)$ and hence
also must neutrino closure techniques similar to FOCUS. Their \qsq{} resolution is nearly identical to that of FOCUS and
they also have an $\approx 65 \%$ negative correlation between $f_+(\qsq)$ bins.
Apart from the two highest BaBar \qsq{} bins, agreement with both FOCUS and 
the LQCD calculations is good.

The values are summarized in Table \ref{alpha} for both $D^0 \rightarrow K^- \ell^+ \nu$ and $D^0 \rightarrow \pi^- \ell^+ \nu$.
The (preliminary) CLEO-c entry in Table \ref{alpha} is based on 281 pb$^{-1}$ of data taken at the $\psi(3770)$ but does not
require a fully reconstructed tagging recoil $\bar D^0$ unlike most CLEO-c $\psi(3770)$ analyzes. This creates a significant
increase in event statistics, but has worse \qsq{} resolution than in CLEO-c fully tagged analyzes. The CLEO-c untagged
analysis still has an order of magnitude better \qsq{} resolution than FOCUS or BaBar. 

The  $K^- \ell^+ \nu$ measurements do not seem terribly consistent between experiments. My naive weighted average of 
the $K^- \ell^+ \nu$ values is $\alpha(K^- \ell^+ \nu) = 0.35 \pm 0.033$ but the CL value that all values are consistent is only
0.9 \%.  The consistency goes up to 39 \% if the preliminary CLEO-c value of $\alpha = 0.19$ is excluded.  My weighted
average of $\alpha(\pi^- \ell^+ \nu) = 0.33 \pm 0.08$. The consistency CL for all three pion measurements is a respectable 56 \%.
 
\begin{table}[htp]
\caption{Modified pole $\alpha$ parameters\label{alpha}}
\begin{center}
\begin{tabular}{|l|c|c|} \hline
& $\alpha (K^- \ell^+ \nu)$ & $\alpha(\pi^- \ell^+ \nu)$ \\ \hline 
CLEO III\cite{cleoqsq} & 0.36  $\pm$  0.10  $\pm$  0.08 & 0.37$^{+0.20}_{-0.31}$  $\pm$  0.15 \\ 
FOCUS\cite{focusqsq} & 0.28  $\pm$  0.08  $\pm$  0.07 &  \\ 
BaBar & 0.43  $\pm$  0.03  $\pm$  0.04 &  \\ 
CLEO-c & 0.19  $\pm$  0.05  $\pm$  0.03 & 0.37  $\pm$  0.09  $\pm$  0.03 \\ 
Belle & 0.52  $\pm$  0.08  $\pm$  0.06 & 0.10  $\pm$  0.21  $\pm$  0.10 \\ \hline
WT AVE   & 0.35 $\pm$ 0.033 & 0.33 $\pm$ 0.08 \\ \hline
\end{tabular}
\end{center}
\end{table}

The data on $\alpha(\pi^- e^+ \nu)$ appears to be consistent with that for $\alpha (K^- e^+ \nu)$
as is the case in LQCD calculations. At this point, the pion data are not sufficiently accurate
to make a really incisive test of the difference between $\alpha(\pi^- e^+ \nu)$ and $\alpha (K^- e^+ \nu)$.

\mysection{Vector $\mathbf{\ell \nu}$ Decays}

Although historically vector decays such as \krzlndk{} have been the most accessible semileptonic
decays in fixed target experiments owing to their ease of isolating a signal, they are
the most complex decays we will discuss. One problem is that a separate form factor is required
for each of the three helicity states of the vector meson. Vector $\ell^+ \nu$ states result in a multihadronic
final state.  For example \krzlndk{} final states can potentially interfere with \kpilndk{} processes
with the $K^- \pi^+$ in various angular momentum waves with each wave requiring its own form factor.
I will concentrate on form factor measurements of Vector $\ell \nu$ decays.
 
\begin{figure}[tbph!]
 \begin{center}
  \includegraphics[width=2.0in]{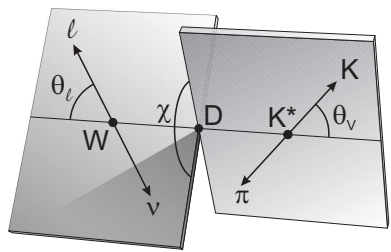}
    \caption{Definition of kinematic variables.
 \label{angles}}
 \end{center}
\end{figure}

\begin{figure}[tbph!]
 \begin{center}
  \includegraphics[width=3.0in]{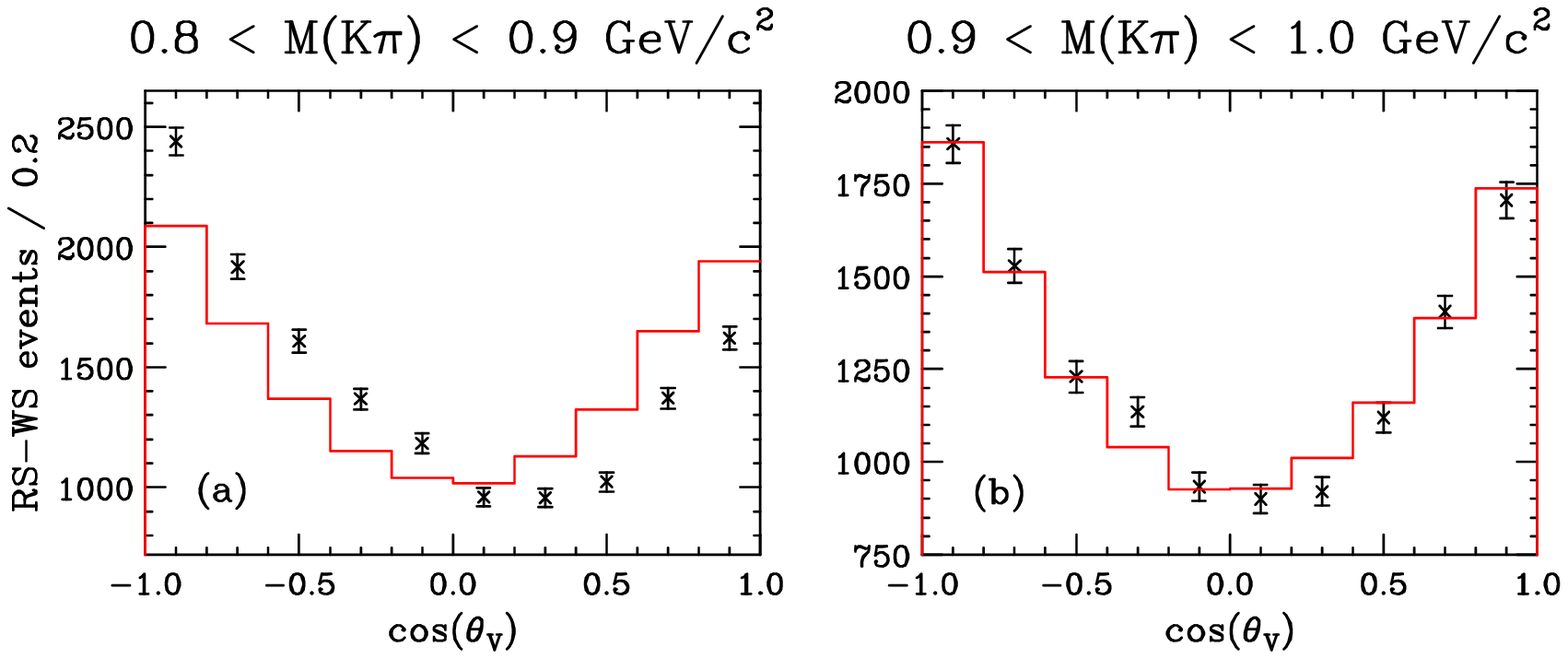}
  \caption{Evidence for s-wave interference in \kpilndk{}.
 \label{asym}}
 \end{center}
\end{figure}

\begin{figure}[tbph!]
 \begin{center}
  \includegraphics[width=3.0in]{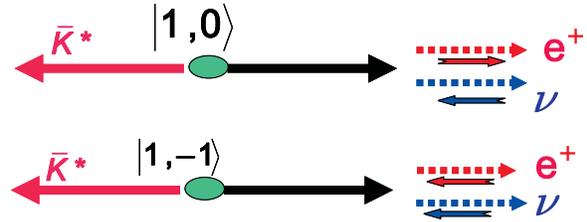}
  \caption{The electron helicity state in the low \qsq{} limit. When the virtual $W^+$ is in the zero
helicity state, the $e^+$ and $\nu$ have the opposite helicity and can be in their charged-current
helicity states. When the virtual $W^+$ is in the $\langle 1 , \pm 1 \rangle$ state the
$e^+$ and $\nu$ must be in the same helicity states and violate the weak helicity rules. 
 \label{hcartoon}}
 \end{center}
\end{figure}

I believe at present, the \kpilndk{} and 
$D_s^+ \rightarrow \phi \ell^+ \nu$ are the only decays with reasonably
well measured form factors.  The three decay angles describing the \kpilndk{}
decay are illustrated in  Fig. \ref{angles}. The other kinematic variables are \qsq{} and \mkpi{}. 
Because the \mkpi{} distribution in \kpilndk{} was an excellent fit to the \krzb{} Breit-Wigner,
it was assumed for many years that any non-resonant component to \kpilndk{} must be negligible.
In 2002, FOCUS observed a strong, forward-backward asymmetry in \costhv{} for events with \mkpi
below the \krzb{} pole with essentially no asymmetry above the pole as shown in Figure \ref{asym}.
The simplest explanation for the \costhv{} asymmetry is an interference between  s-wave and p-wave
amplitudes creating a linear \costhv{} term. The phase of the s-wave amplitude must be such that 
its phase is nearly orthogonal with the Breit-Wigner ($BW$) phase for $\mkpi{} > m(\krzb)$. 
The (acoplanarity) averaged $|\mathcal{A}|^2$  in the zero lepton mass limit (Eq. (\ref{KSE})) is constructed from the Breit-Wigner ($BW$), s-wave
amplitude ($Ae^{ - i\delta }$), and the helicity basis form factors \Hpls{}, \Hmin{}, \Hzer{} that describe
the $W^+$ coupling to each of the \krzb{} spin states~\cite{KS}.  We also need an additional form factor (\hzer{})
describing the coupling to the s-wave amplitude.  
\begin{widetext}
\begin{eqnarray}
\int {\left| {\rm{A}} \right|^2 d\chi }  = \frac{1}{8}q^2 \left\{ \begin{array}{l}
 \left( {(1 + \cos \theta _l )\sin \theta _V } \right)^2 \left| {H_ +  (q^2 )} \right|^2 \left| {BW} \right|^2  \\ 
  + \left( {(1 - \cos \theta _l )\sin \theta _V } \right)^2 \left| {H_ -  (q^2 )} \right|^2 \left| {BW} \right|^2  \\ 
  + \left( {2\sin \theta _l \cos \theta _V } \right)^2 \left| {H_0 (q^2 )} \right|^2 \left| {BW} \right|^2  \\ 
  + 8\left( {\sin ^2 \theta _l \cos \theta _V } \right)H_0 (q^2 )h_o (q^2 ){\mathop{\rm Re}\nolimits} \left\{ {Ae^{ - i\delta } BW} \right\} \\ 
  + O(A^2 ). \\ 
 \end{array} \right\}
\label{KSE}
\end{eqnarray}
The   \Hpls{}, \Hmin{}, and \Hzer{} form factors are linear combinations of two axial and one vector form factor as indicated
in Eq. (\ref{helicity}). 
\begin{eqnarray}
H_\pm(\qsq) &=&
   (M_D+\mkpi)A_1(\qsq)\mp 2{M_D K\over M_D+m_{K\pi}}V(\qsq) \,,
                               \nonumber \\
H_0(\qsq) &=&
   {1\over 2\mkpi\sqrt{\qsq}}
   \left[
    (M^2_D -m^2_{K\pi}-\qsq)(M_D+\mkpi)A_1(\qsq)          
    -4{M^2_D K^2\over M_D+\mkpi}A_2(\qsq) \right]. \label{helicity} \nonumber \\ 
\label{helform}
\end{eqnarray}
\end{widetext}
Eq. (\ref{helform}) shows that as $\qsq{} \rightarrow 0$, both \Hpls{} and \Hmin{} approach a constant. 
Since the helicity intensity contributions are proportional to $\qsq{} H^2_\pm(\qsq{})$  ( Eq. (\ref{KSE}))
the $H_\pm$ intensity contributions vanish in this limit. Figure \ref{hcartoon}
explains why this is true. As $\qsq{} \rightarrow 0$, the $e^+$ and $\nu$ become collinear with the virtual $W^+$.
For \Hpls{} and \Hmin{}, the virtual $W^+$ must be in either the $| 1 , \pm 1 \rangle$ state which means
that the  $e^+$ and $\nu$ must both appear as either righthanded or lefthanded thus violating the charged current helicity rules. 
Hence $\qsq{}H_\pm (\qsq{})$ vanishes at low \qsq{}.  For \Hzer{}, the $W^+$ is in $| 1 , 0 \rangle$ state thus allowing the 
$e^+$ and $\nu$ to be in their (opposite) natural helicity state.  Hence at low \qsq{},  $\Hzer{} \rightarrow 1/\sqrt{\qsq{}}$
which allows for  \krzmndk{} decays as $\qsq{}\rightarrow 0$. Presumably $\hzer{} \rightarrow 1/\sqrt{\qsq{}}$ as well since
it also describes a process with $W^+$ is in $| 1 , 0 \rangle$ state 
  
Vector $\ell^+ \nu$ processes have been traditionally analyzed using a spectroscopic pole dominance 
model for 
$V(\qsq{})$, $A_1 (\qsq{})$ and $A_2(\qsq)$. The vector pole is at the mass of the $D_s^*$  ; while both axial poles are set to 2.5 GeV.  

\begin{figure}[tbph!]
 \begin{center}
  \includegraphics[width=3.0in]{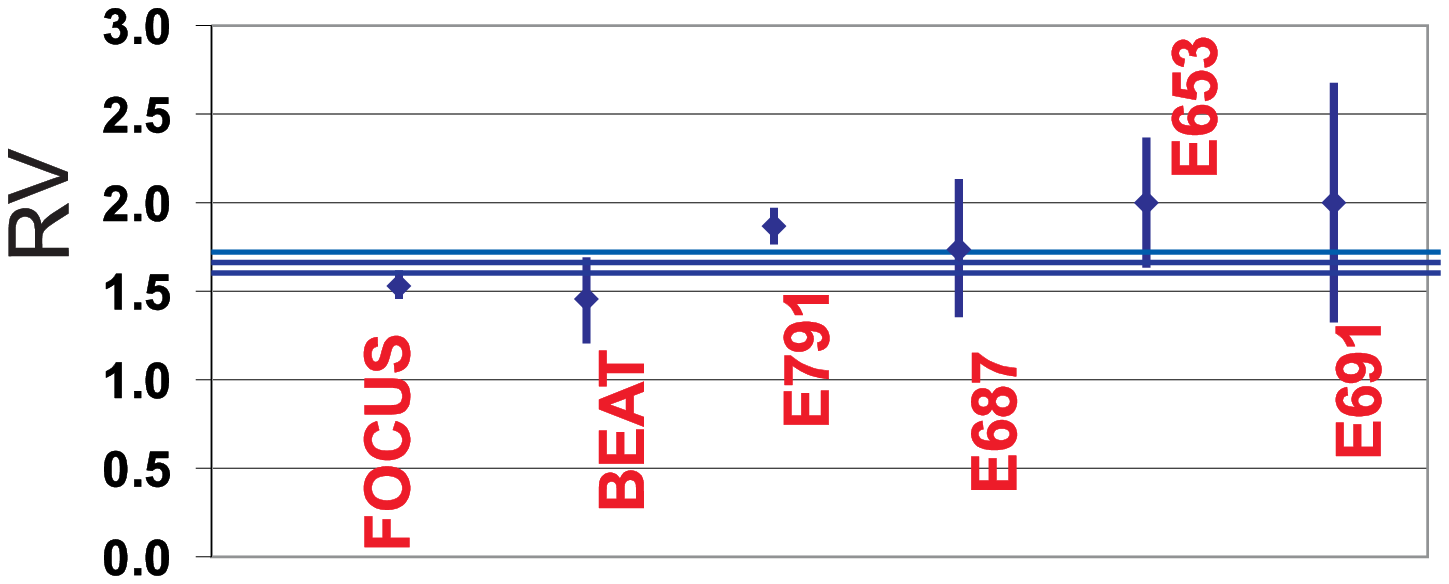}
 \includegraphics[width=3.5in]{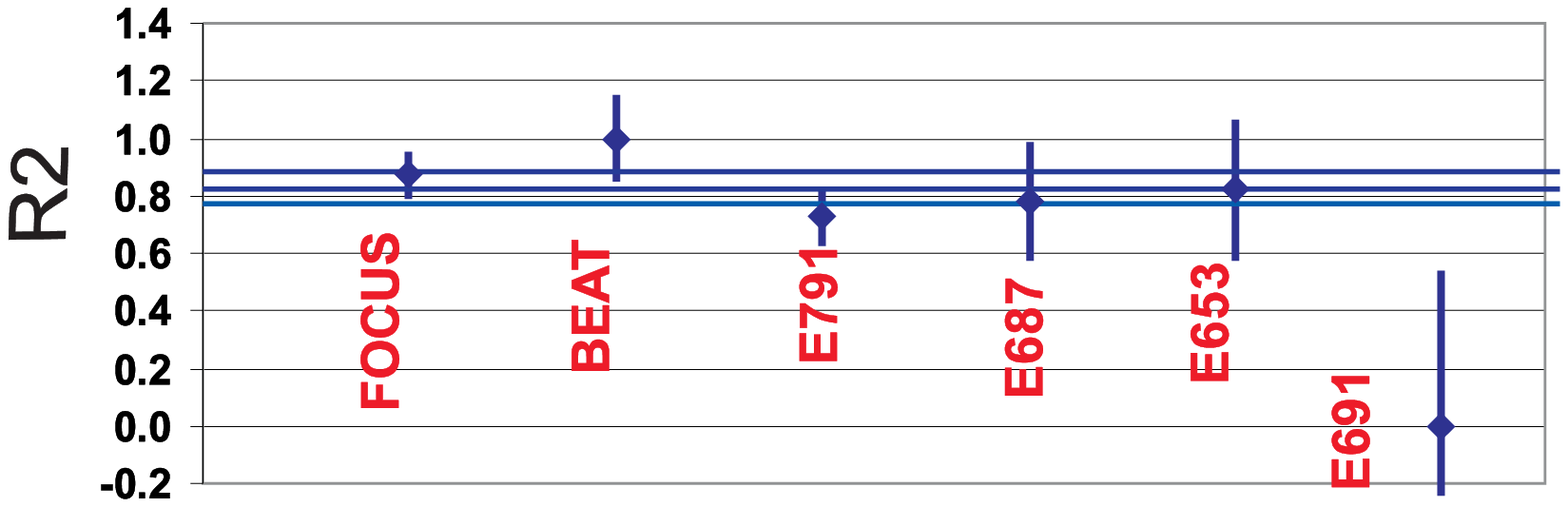}
  \caption{The \rvee and \rtwo form factor ratios for \krzlndk{} measured by six experiments. The blue
lines are the weighted average of all six measurements.
 \label{kpilnu}}
 \end{center}
\end{figure}

Under these assumptions the shape of the \krzmndk{} intensity (apart from the s-wave effect) is fully determined
from the ratio of the axial and vector form factors at \qsq{} = 0.  Traditionally the variables are $\rvee = V(0)/A_1(0)$ and
$\rtwo = V(0)/A_1(0)$.
A long series of measurements has been made for \krzlndk{} and \philndk{} over the years under the assumption with spectroscopic pole dominance~\cite{oldformfactor}. A wide
range of theoretical techniques have been employed to predict the form factor ratios more or 
less successfully~\cite{quark}\cite{lqcd}\cite{ball}.
Figures \ref{kpilnu} summarize these measurements.  My weighted average is $\rvee= 1.618 \pm 0.055$ and  
$\rtwo = 0.830 \pm 0.054$ with a confidence level of 6.7\% that all \rvee{} values are consistent and 42\% that all \rtwo{} values are consistent.  
Only the latest measurement by FOCUS includes the s-wave contribution-- including it with the ad-hoc assumption that
the \hzer{} form factor for the $K \pi$ s-wave contribution is the same as the \Hzer{} form factor for the zero helicity \krzb contribution. 

\begin{figure}[tbph!]
 \begin{center}
  \includegraphics[width=2.7in]{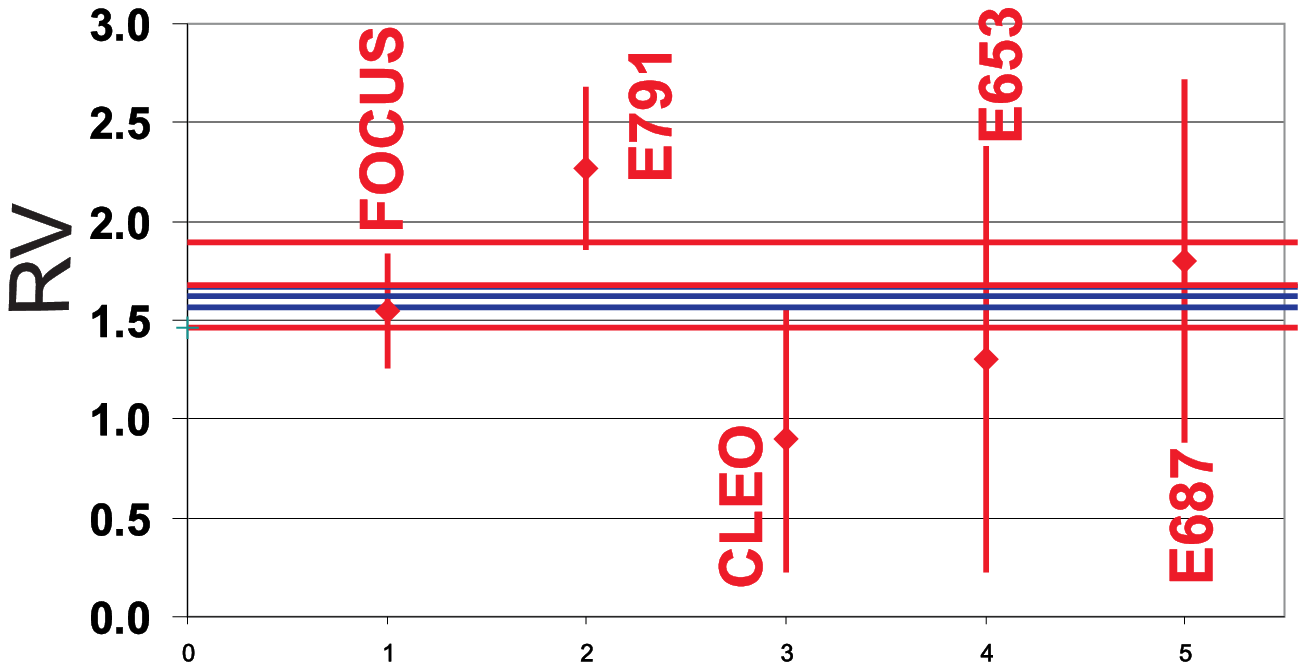} \vskip .2in 
  \includegraphics[width=2.7in]{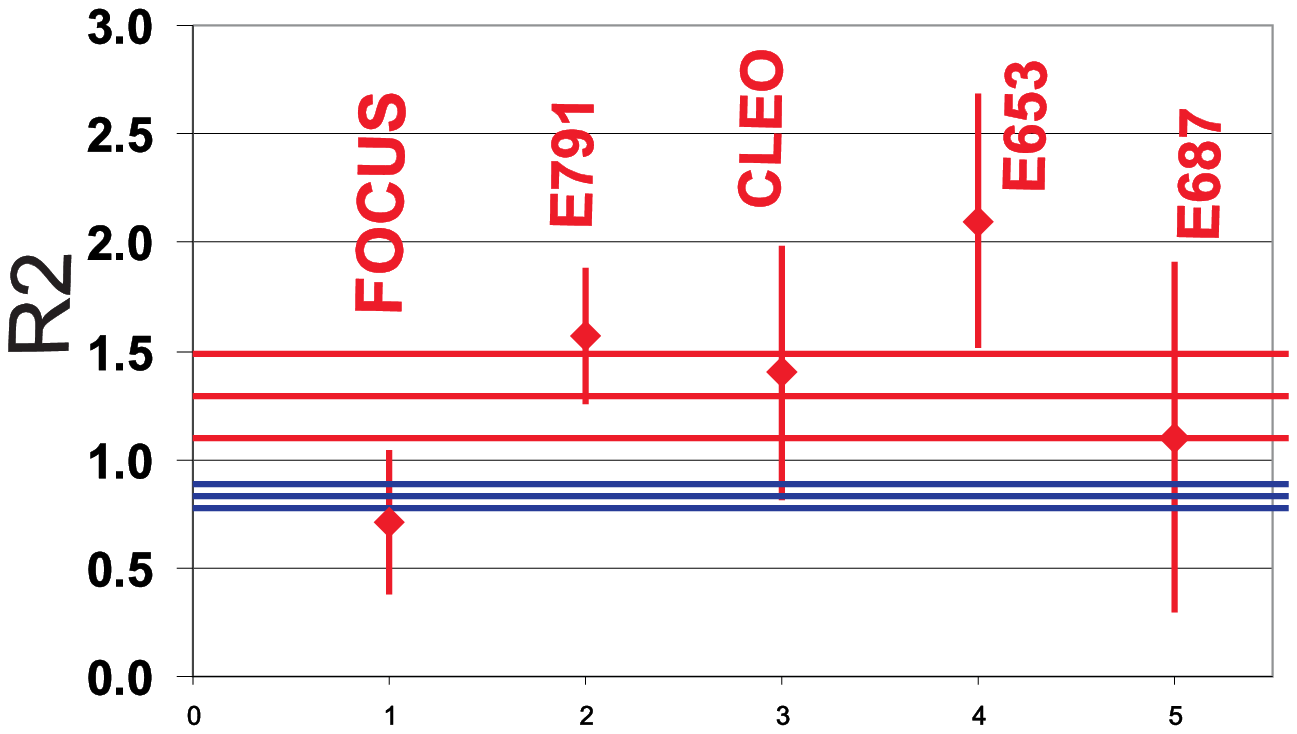}
  \caption{The \rvee and \rtwo form factor ratios for \philndk{} measured by five experiments. The blue
lines are the weighted average of the \krzlndk{} form factors shown in Fig. \ref{kpilnu}. It was
expected that the \philndk{} form factors should be consistent with the \krzlndk{} form factors.
 \label{philnu}}
 \end{center}
\end{figure}

The experimental situation with \philndk{} shown in Fig. \ref{philnu}
is somewhat less clear~\cite{phigoof}. By SU(3) symmetry and explicit calculation, the \rvee{} and \rtwo{} form factor ratios for \krzlndk{} and \philndk{} decays are
expected to lie within $\approx 10\%$ of each other~\cite{russian}. This is true for \rvee{}, but previous to the very recent measurement by the 
FOCUS~\cite{focusphi}, \rtwo{} for  \philndk{} was roughly a factor of two larger than that
for \krzlndk{} although there is a 27\% confidence level that all published \rtwo{} values for \philndk{}
are consistent.

Given the failure of the spectroscopic pole model in 
pseudoscalar $\ell^+ \nu$ decays,  and the fact that the $\qsq{}_{\rm max}$ for \krzlndk{} is 
even further from the $D_s^{+*}$ pole than the case for $K^- \ell+ \nu$ it seems unlikely that spectroscopic pole dominance is a good model for
axial and vector form factors relevant to vector $\ell^+ \nu$ decay. 
Although most groups reporting \rvee{} and \rtwo{} values show that
their fits roughly reproduce the various \costhv, \costhl, \qsq{}, 
and $\chi$ projections observed in their data,
there have been no quantitative tests to my knowledge on the validity of the 
the spectroscopic pole assumptions in vector $\ell^+ \nu$ charm decay. 
Fajfer and Kamenik \cite{FK} have proposed an effective pole descriptions of
the vector and two axial form factors used in Eq. (\ref{KSE}).  For example their $V(\qsq)$ parameterization 
is identical to the $f_+(\qsq{})$ given in Eq. (\ref{modpole}).  But I know of no attempts
to fit for either the effective pole parameters of Fajfer and Kamenik  or
simple effective poles such as those displayed in Fig. \ref{mpole} for $K^- \ell^+ \nu$.
The problem is that the spectroscopic pole constraint is such a powerful 
constraint that releasing it would severely inflate errors on \rvee{} and \rtwo{}. 

As a first attempt to study \krzlndk{} free from the constraining assumption of 
spectroscopic pole dominance,
FOCUS \cite{focus-helicity} developed a non-parametric method for studying
the helicity basis form factors. As shown in Eq. (\ref{KSE}), after integration
by the acoplanarity $\chi$ to kill interference between different helicity states, 
the decay intensity greatly simplifies into a sum of just four terms
proportional to: \Hszer{}, \Hspls{}, \Hmin{}, and \Hint{}. 
Each term is associated with a unique angular distribution which can be used to project out each individual term. 
The projection can be done by making  4 weighted histograms using projective weights based on the
\costhv{} and \costhl{} for each event. 

\begin{figure}[tbph!]
 \begin{center}
  \includegraphics[width=3.in]{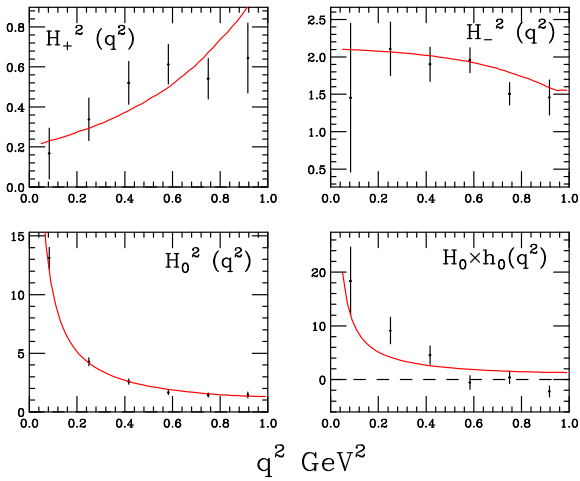}
  \caption{The four helicity form factor products obtained in a preliminary analysis using 281 pb$^{-1}$ data set
from CLEO.  The curves represent the model of Reference \cite{formfactor}.
 \label{hsq}}
 \end{center}
\end{figure}

Figure \ref{hsq} shows the four weighted histograms from a preliminary analysis of 281 pb$^{-1}$ of $\psi(3770)$ 
CLEO data.  The CLEO data are far superior for this analysis because of its nearly order of magnitude
better \qsq{} resolution than the resolution in  a fixed target experiment such as FOCUS.  
 
Figure \ref{hsq} shows the expected behavior that $H^2_\pm(\qsq{}) \rightarrow {\rm constant}$ as
$\qsq{} \rightarrow 0$ while \Hszer{} and \Hint{} approaches $1/\qsq{}$.  The curves give the
helicity form factors according to Eq. (\ref{KSE}), using spectroscopic pole dominance and 
the \rvee{}, \rtwo{}, and s-wave parameters measured by FOCUS. Apart from the \Hint{} form 
factor product the spectroscopic pole dominance model is a fairly good match to the 
CLEO non-parametric analysis.  This suggests that the ad-hoc assumption that \hzer{}=\Hzer{}
is questionable but it will probably take more data to gain insight into the nature of the 
discrepancy.

\begin{figure}[tbph!]
 \begin{center}
  \includegraphics[width=3.in]{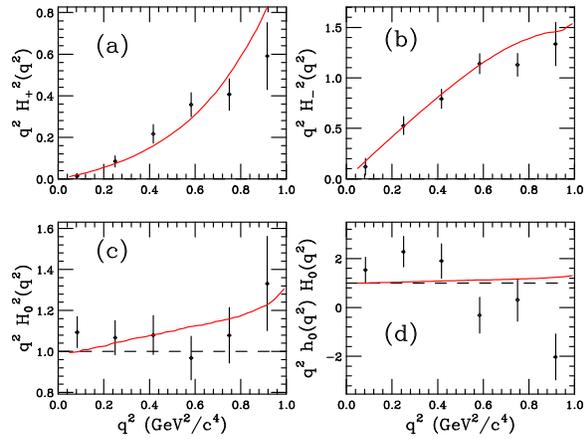}
  \caption{Non-parametric form factor products obtained for the data sample
(multiplied by \qsq{}) The reconstructed form factor products are shown as the points with error bars,
where the error bars represent the statistical uncertainties.
The solid curves in the histograms represent a form factor model described
in Ref.~\cite{formfactor}. 
The histogram plots are:
(a)~$\qsq{} H_+^2(\qsq)$,
(b)~$\qsq{} H_-^2(\qsq)$,
(c)~$\qsq{} H_0^2(\qsq)$, and
(d)~$\qsq{} h_0(\qsq) H_0(\qsq)$.  The form factors are normalized such that $\qsq{} h_0(\qsq) H_0(\qsq) \rightarrow 1$
as $\qsq{} \rightarrow 0$.
 \label{qsqhsq}}
 \end{center}
\end{figure}

Figure \ref{qsqhsq} gives a different insight into the helicity basis form factors by
plotting the intensity contributions of each of the form factor products. This is the
form factor product multiplied by \qsq{}. 
Since \qsq{}\Hszer{} is nearly constant, we normalized form factors such that $\qsq{}\Hszer{} = 1$ at \qsq{} = 0.
As one can see from Eq. (\ref {KSE}), both \qsq{}\Hspls{} and \qsq{}\Hsmin{} rise from zero with increasing \qsq{} until they both equal \qsq{}\Hszer{} at  \qsq{}$_{max}$.
As \qsq{} is increased from 0 the \krzlndk{} $\theta_V$ distribution evolves from $\costhv^2$ (100 \% longitudinally polarized) to a flat distribution
(unpolarized).  

\begin{figure}[tbph!]
 \begin{center}
  \includegraphics[width=3.in]{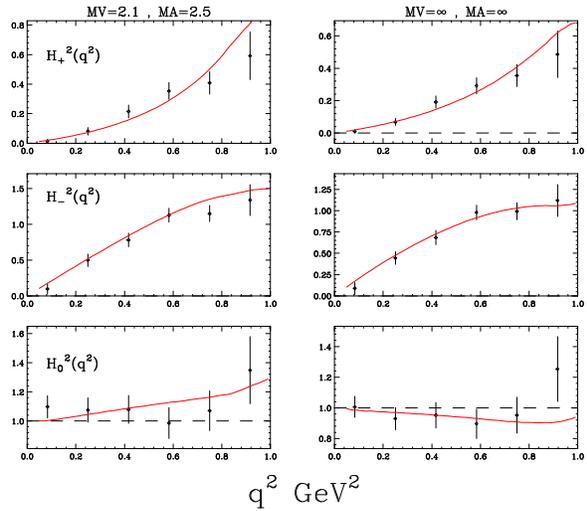}
  \caption{Non-parametric form factor products obtained for data (multiplied by \qsq{})  
The solid curves are based on the $s$-wave model and measurements described in Reference~\cite{formfactor} 
The reconstructed form factor products are the points with error bars. The three plots on the right are the usual model with the 
spectroscopic pole masses; while the three plots on the right are run with all pole masses with the axial and vector
pole masses taken to infinity
 \label{infpole}}
 \end{center}
\end{figure}

What can we learn about the pole masses?  Unfortunately Fig. \ref{infpole} shows that the present
data are insufficient to learn anything useful about the pole masses.  On the 
left of Figure \ref{infpole} the helicity form factors are compared to 
a model generated with the FOCUS form factor ratios and the standard pole masses
of 2.1 GeV for the vector pole and 2.5 GeV for the two axial poles. On the right side
of Fig. \ref{infpole} the form factors are compared to a model where the pole masses
are set to infinity.  Both models fit the data equally well.

\begin{figure}[tbph!]
 \begin{center}
  \includegraphics[width=3.in]{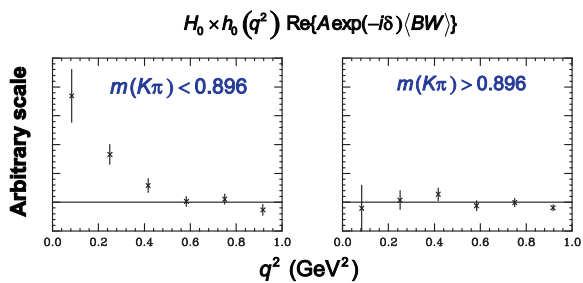}
  \caption{The s-wave interference term for events below the \krzb{} pole (left) and above the pole (right).
The interference term depends on the s-wave phase relative to the phase average phase of each 
half of the Breit-Wigner.  All of the \costhv{} interference observed by FOCUS was also below the 
\krzb pole as shown in Fig. \ref{asym}
 \label{split}}
 \end{center}
\end{figure}

What can we learn about the phase of the s-wave contribution?  Recall in 
Figure \ref{asym} the \costhv{} asymmetry created by the interference between the
s-wave and \krzlndk{} only appeared below the \krzb pole in FOCUS data and meaning
that the s-wave phase was orthogonal with the $\mkpi > m(\krzb)$ half
of the  Breit-Wigner amplitude.  As Figure \ref{split} shows, the same thing happens in CLEO data :
the effective \Hint{} disappears above the \krzb pole and is very strong
below the pole.  The amplitude $A$ of the s-wave piece is arbitrary since
using interference we can only observe the product $A~h_0 (\qsq{})$.  This
means any change in $A$ scale can be compensated by a change of scale in $h_0 (\qsq{})$.
The fact that the \Hint{} data was a tolerable match (at least in the low \qsq{} region) to the FOCUS curve in Figure \ref{hsq} 
does imply, however, that the s-wave amplitude observed in CLEO is consistent with that of FOCUS.  A more formal
fit of the s-wave parameters is in progress.

Finally, is there evidence for higher $K^- \pi^+$ angular momentum amplitudes in \kpilndk{}? 
We searched for possible additional interference terms such as a $d$-wave contribution:
\begin{widetext}
$$4\,\sinthlsq~\costhv~(3\,\cos^2 \thv - 1)\,H_0(q^2)\,h^{(d)}_0(q^2)\,
{\mathop{\mathrm{Re}}\nolimits}\{\mathrm{A}_de^{-i\delta_d} \bw\}$$  
or an $f$-wave contribution: 
$$4\,\sinthlsq~\costhv~ (5\,\cos^3 \thv - 3 \cos \theta)\,H_0(q^2)\,h^{(f)}_0(q^2)\,
{\mathop{\mathrm{Re}}\nolimits}\{A_f e^{-i\delta_f} \bw \}.$$ 
\end{widetext}
As shown in Figure \ref{fdwave} there is no evidence for such additional contributions:
\begin{figure}[tbph!]
 \begin{center}
  \includegraphics[width=3.in]{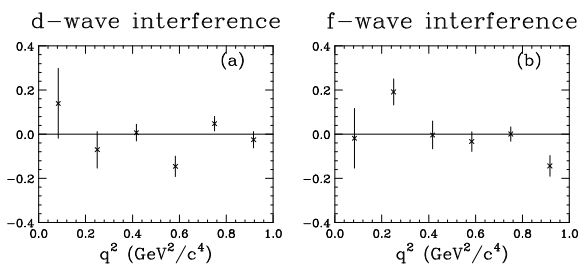}
  \caption{Search for (a) $d$-wave and (b) $f$-wave interference effects as described in the text.
 \label{fdwave}}
 \end{center}
\end{figure}

\mysection{Summary}

A great deal of progress has been made in charm semileptonic decay in the last few 
years.  A new set of precision semileptonic branching ratios have been made 
available from CLEO.  These include both exclusive mesonic branching fractions
as well as inclusive semileptonic branching fractions for the $D^0$ and $D^+$.
This data suggests that the known exclusive decays comes close to
saturating the measured inclusive branching fraction, and that the
inclusive semileptonic widths for the $D^+$ and $D^0$ are equal as expected.

The first precision measurements of charm fully leptonic decay have been
made by CLEO ($D^+ \rightarrow \mu^+ \nu$) and BaBar ($D_s^+ \rightarrow \mu^+ \nu$).
Both experiments produce $\approx 8\%$ measurements of the meson decay constants ($f_D$) that
are consistent with LQCD calculations and with comparable uncertainty to the calculations.

Several new precision, non-parametric measurements have been made of the $f_+(\qsq{})$ form factor in 
$D^0 \rightarrow K^- \ell^+ \nu$.  At present the situation is a bit murky. The earlier
measurements, tend to agree with each other as well as the LQCD calculations on the form factor shape.
One of the new preliminary measurement has a significantly different shape parameter $\alpha$.

Finally progress in understanding vector $\ell^+ \nu$ decays was reviewed.  These have
historically been analyzed under the assumption of spectroscopic pole dominance.  Experiments
have obtained consistent results under this assumption, but as of yet there have been
no incisive tests of spectroscopic pole dominance.   We concluded by describing a first, preliminary
non-parametric look at the \kpilndk{} form factors.  Although the results were very consistent
with the traditional pole dominance fits, the data was not precise enough to incisively
measure \qsq{} dependence of the axial and vector form factors and thus test spectroscopic dominance.  This
preliminary analysis confirms the existence of an $s$-wave effect first observed by FOCUS \cite{swave},
and was unable to obtain evidence for $d$ and $f$-waves.

\end{document}